\documentclass[useAMS,usenatbib]{mn2e}
\usepackage[utf8x]{inputenc}
\usepackage{epsfig}
\usepackage{graphicx}
\title[Emission line tomography of CC Scl and V2051 Oph]{Emission line tomography of the short period cataclysmic variables CC Scl and V2051 Oph}
\author[P. Longa-Pe\~na, D. Steeghs and T. Marsh]{P. Longa-Pe\~na, D. Steeghs and T. Marsh\\
$^{}$Department of Physics, University of Warwick, Coventry CV4 7AL, UK}

\begin{document}

\maketitle

\label{firstpage}

\begin{abstract}

    We present time-series spectroscopy of two short period cataclysmic variables, CC Scl and V2051 Oph, to test the efficiency of Doppler tomography-based methods in constraining orbital parameters of evolved cataclysmic variables. We find that the Ca~II triplet lines offer superior diagnostics, revealing emission components from the mass donors and sharp images of the accretion discs.  Furthermore, we use Monte-Carlo methods to estimate the uncertainties from ensembles of Doppler maps.  We compare our new methods against traditional radial velocity methods and show that they offer a valid route towards system parameter determination. 
    Our analysis of CC Scl suggests a low mass ratio of $q=0.08\pm0.03$ with a primary velocity of $K_1=37\pm14$~km/s. This mass ratio is in between the pre- and post-period minimum status, however our $K_1$ solution favours a post-period minimum system.  Our derived parameters for V2051 Oph ($q= 0.16\pm 0.03$, $K_1=97\pm10$~km/s) are in agreement with the  eclipse solution ($q=0.19\pm0.03$), offering a direct validation of our methods.

\end{abstract}

\section{Introduction}

Accretion onto compact objects powers a diverse set of high-energy events and source types, and is linked to our understanding of Type Ia supernovae (\citealt{1973ApJ...186.1007W}, \citealt{1986ApJ...301..601W}, \citealt{2004Natur.431.1069R}), gravitational wave emission events (\citealt{1992Natur.355..614C}) and gamma-ray bursts (\citealt{1986ApJ...308L..43P}, \citealt{2005Natur.438..988B}). To understand accretion, we need to study populations of compact binary systems of different type. The cataclysmic variables (CVs) are compact binaries consisting of an accreting white dwarf (WD) primary star and a secondary star that fills its Roche lobe (\citealt{Warner1995}). Their prevalence makes CV populations particularly useful to further our understanding of accretion and compact binary evolution.

It is believed that a large percentage of CVs should have evolved towards short orbital periods ($\sim 80$ mins). At this so-called period minimum, the mass of the donor becomes so low that it becomes degenerate and no longer shrinks, but expands in response to mass loss, and the period evolution reverses sign, leading to a period increase. Thus a CV is expected to spend a significant fraction of its time with an orbital period near the period minimum. Because of this, a significant accumulation of CVs at the minimum period is expected. This accumulation is often called the `period minimum spike' (e.g. \citealt{1999MNRAS.309.1034K}). This prediction, however, has only recently been firmly supported by observations, largely thanks to  deep survey samples such as provided by the Sloan Digital Sky Survey (SDSS) (\citealt{2009MNRAS.397.2170G}).   While the standard CV evolution model can explain the bulk features of the CV period distribution, more quantitative comparisons between evolutionary models and the CV population is desired, in particular for those systems where more than just the binary period is known.
Theoretical models of the evolution of CVs predict the distribution of their orbital parameters: orbital periods, mass ratio $q=M_2/M_1$ and the individual masses of the two stars, $(M_1,M_2)$.  Regrettably all too often, the only reliable physical parameter that can be measured for most CVs is their orbital period.   In order to validate the theory and pinpoint the evolutionary track of CVs -- evolving towards (pre-bounce) or  from the period minimum (post-bounce)-- more information about the parameters of individual systems is needed.  Particularly, the mass of the secondary star can differentiate a pre- from a post-bouncer as the evolution of CVs is controlled by the properties of their secondaries.

The optical spectra of systems with periods close to the period minimum are often dominated by emission from the white dwarf and the accretion disc, but typically do not show discernible signatures of the faint, low mass secondary stars. The presence of disc emission lines offers some insight into the WD radial velocity.  The most widely used technique to analyse them is the double Gaussian of \cite{1980ApJ...238..946S} that uses the wings of the emission lines to determine the binary motion of the disc and thus the WD if the disc is not distorted. This method is based on the fact that any asymmetry will be confined to the outer region of the disc, while the wings trace gas close to the WD. In practice, however, this is not always true.    Non-axi-symmetric components such as the bright spot can distort the derived radial velocity curve out to large velocities from line centre.  To mitigate this, the diagnostic diagram of \cite*{1986ApJ...308..765S} is often used, although even this  method is prone to systematics, but it can at least be used to estimate the significance of such asymmetries.

Without a detection of the secondary star or a fortuitous inclination that leads to eclipses, it is thus quite difficult to derive robust binary parameters for many short period CVs. The until-recently-neglected Ca~II triplet, however, has proven to be a promising candidate for orbital parameter constraints as shown by \cite{2010MNRAS.401.1857V}.  Despite the relative weakness of the Ca~II triplet with respect to the Balmer lines, it is commonly found in short period CVs and can track the information of the secondary star better than the Balmer lines due to its lower ionisation energy.

In this paper, we present new methods to derive dynamical constraints in short period CVs.  We present time-resolved  spectra of CC Scl and V2051 Oph (Section \ref{back} and \ref{spectrum}) to test these methods. In Section \ref{ccscl_dt} we derive dynamical constraints on the radial velocity of each system to estimate their mass ratios. % We present a method that uses Doppler maps of the Ca~II $8662.14$~\AA\ to estimate the value of the phase zero $\Phi_0$, the primary radial velocity $K_1$, and the radial velocity of the irradiated face  of the donor $K_{em}$,  and hence to constrain the value of the mass ratio by calculating a K-correction, the difference of the radial velocity of the irradiated face of the donor and its actual centre of mass ($\Delta K=K_2-K_{em}$).  Furthermore, we propose a variation from the Doppler tomography secondary emission technique to constrain the value of the systemic velocity $\gamma$ (e.g. \citealt{2003ApJ...590.1041C}), automating the method and provide uncertainties to all our calculated parameters by means of the bootstrapping technique. 
In Section \ref{rvelanal}, we compare the traditional methods of radial velocity determination and compare the performance of these methods for different lines against the Doppler map-based techniques. Finally, in Section \ref{discussions} we analyse the efficiency of the method and present our findings for the two CV systems.

\section{Observations and Reduction}

The observations were taken in July 2010, using the MagE (Magellan Echellette; \citealt{2008SPIE.7014E..54M}) spectrograph mounted on the Magellan-Clay  telescope at Las Campanas Observatory, La Serena, Chile.  We used the 1" slit  with the 175 lines/mm grating covering $\sim3100-11200$~\AA\ at a resolution of R=4100 ($\sim71$~km/s).  We obtained 26  spectra of CC Scl on the 8th and 43 of V2051 Oph on the 9th with 180~[s] exposure time respectively. The MagE detector is a 2048x1024 pixel E2V CCD with 35 micron pixels and was readout unbinned in slow readout mode.

The spectra  were reduced with the \texttt{Python} based pipeline written by Dan Kelson\footnote{ Carnegie Observatories Software Repository: http://code.obs.carnegiescience.edu/mage-pipeline} which performs typical calibrations: flat-fielding, sky background subtraction, optimal extraction and wavelength calibration.  The wavelength calibrations were derived from regular Thorium-Argon lamp exposures taken during the respective nights. These provide suitable lines over the entire wavelength range.  The pipeline selects the closest lamp exposures in time to each science spectrum.

Raw spectra were flux calibrated using the spectrum of the flux standard Feige 110, observed at the end of each night using both a wide slit and the same slit width used for the target exposures. The wide-slit observation was used for determining the wavelength dependent flux calibration. The spectra were then divided by the ratio between the narrow and wide slit spectra in order to apply a 1st-order slit loss correction.

Due to the poor signal to noise ratio of the first ($\sim9400-11200$~ \AA) and last orders($\sim3100-3500$ ~\AA), these were not included in the analysis.  We did not attempt to remove telluric features from the spectra. 

% 
% \begin{table}
%  \begin{tabular}{c c c c c}
% \hline
% \hline
% Object&Date&Phase  &Number of&Exp time\\
%  & &Coverage&Spectra& (s)\\
% \hline
% 
% CC Scl & $08/06/2010$ &$ 1.1$&$26$ & $180$ \\
% \hline
% V2051 & $09/06/2010$&$2.3$&$43$& $180$ \\
%  Oph & & & & \\
% \hline
%  \end{tabular}
% \caption{MagE observations details.}
% \label{tab:tabla_obs}
% \end{table}

\section{Results}

\subsection{Background}
\label{back}

CC Scl was independently discovered as the ROSAT source 1RXS J2315532.3-304855 \citep{2000AN....321....1S} and in the Edinburgh-Cape Blue Object Survey as EC 23128-3105 \citep{2001MNRAS.325...89C}.  From time resolved spectroscopy, Chen et al. derived an orbital period of 0.0584(2)d, $K_1=35\pm10$~km/s and $0.06<$q$<0.09$.  They also reported a possibly transient feature resembling an eclipse present in some of their photometric light curves.  This feature was also reported by \cite*{2004MNRAS.354..321T}.

\cite{2001IBVS.5023....1I} presented a significant amount of photometry covering a six-month baseline, finding superhumps and classifying CC Scl as a SU UMa type dwarf nova. They claimed a super hump period of 0.078 d, which would be $30\%$ larger than the orbital period.  This would be the largest fractional superhump excess in a SU UMa star followed by $7.7\%$ observed in TU Men.
During the November 2011 outburst,  \cite{2012MNRAS.427.1004W} attempted to refine the orbital period photometrically, but this was not consistent with the spectroscopic period.  \citeauthor{2012MNRAS.427.1004W} also found a periodicity of 389.4~s, identified in the optical, UV and X-ray spectra. They associated this perdiodicity with the white dwarf's spin period and classified CC Scl as a super-humping intermediate polar below the period gap.  Their data resolution did not allow them to observe any shallow eclipse.

Recently, \cite{vsnet} have confirmed CC Scl as an eclipsing system.  They presented a light curve showing a short and shallow eclipse, lasting $\sim0.08$ orbital phases, that can be used to further refine the ephemeris and orbital period of the system to $(BJD) = 2456668.00638(9) + 0.058567233(8) E$.  This latest period is also not consistent with the period proposed by \cite{2012MNRAS.427.1004W}.

V2051 Oph was discovered by  \cite{1972IBVS..663....1S}. It is classified as an eclipsing SU UMa type CV. \cite{1998IBVS.4644....1K} confirmed this classification, using the photometric detection of superhumps in the light curve of V2051 Oph during outburst.

 \cite{1998MNRAS.300..233B} used ground-based high-speed eclipse photometry and HST spectroscopy to derive the binary geometry and to estimate the masses.  They reported the mass ratio to be $q=0.19\pm0.03$, an inclination  $i=83^o\pm2^o$ and  masses of $M_1=0.78\pm0.06M_{\odot}$ and $M_2=0.15\pm0.03M_{\odot}$.  Their photometric model results in a projected radial velocity for the white dwarf of $K_1=83\pm12$~km/s while the secondary moves at $K_2=436\pm11$~km/s. The $K_1$ value is in excellent agreement with the radial velocity analysis of \cite{2008A&A...487..611P}, who find $K_1=85\pm12$~km/s.   

V2051 Oph shows asymmetry in its  accretion disc, seen in the eclipse mapping of  \cite{2006AJ....131.2185S} and \cite{2007AJ....134..867B}.  Asymmetry was also seen in the spectra of \cite{2001MNRAS.323..484S}.

\subsection{Spectrum}
\label{spectrum}

\begin{figure*}

 \includegraphics[width=1\linewidth, trim= 0cm 1.4cm 0cm 0cm, clip=true]{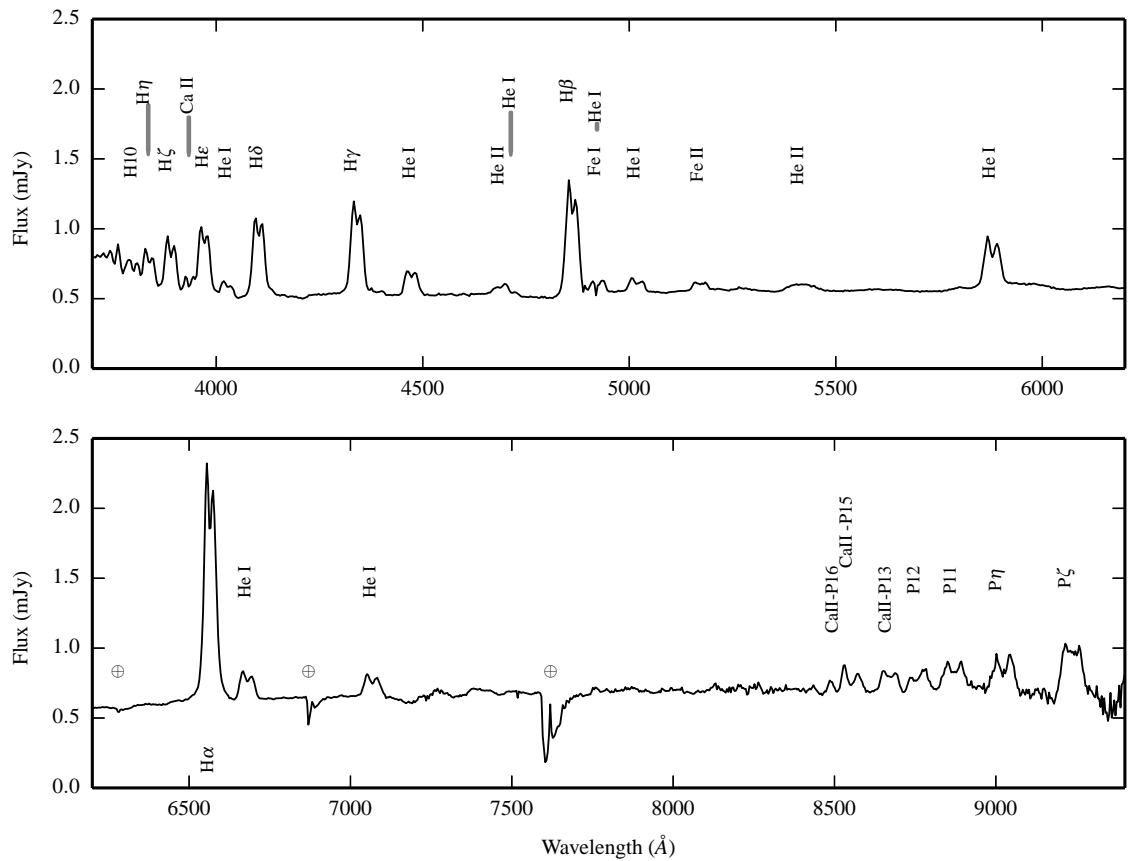}
\caption{Average spectrum of CC Scl.  Each of the 26 spectra is given equal weighting and no correction for orbital motion has been made.}
\label{average_ccscl}
\end{figure*}
In the case of V2051 Oph, we found features unnoticed until now, such as the blend between  N~III $4640.64$~\AA, HeI $4713.20$~\AA\ and He~II $4686.75$~\AA.  N~III $4640.64$~\AA  is part of the Bowen blend, which also includes some C~III components. In X-ray binaries, this blend often contains strong components from the irradiated face of the donor (Bowen blend fluorescence, see for example \citealt{2002ApJ...568..273S}, \citealt{2006MNRAS.373.1235C} and \citealt{2003ApJ...590.1041C}).  Despite our high resolution, the S/N of the blend was too low to attempt to find the secondary by this method.  Si~II and O~II in emission, and O~I in absorption are also present in our spectra.

\begin{figure*}

 \includegraphics[width=0.95\linewidth, trim= 0cm 1.4cm 0cm 0cm, clip=true]{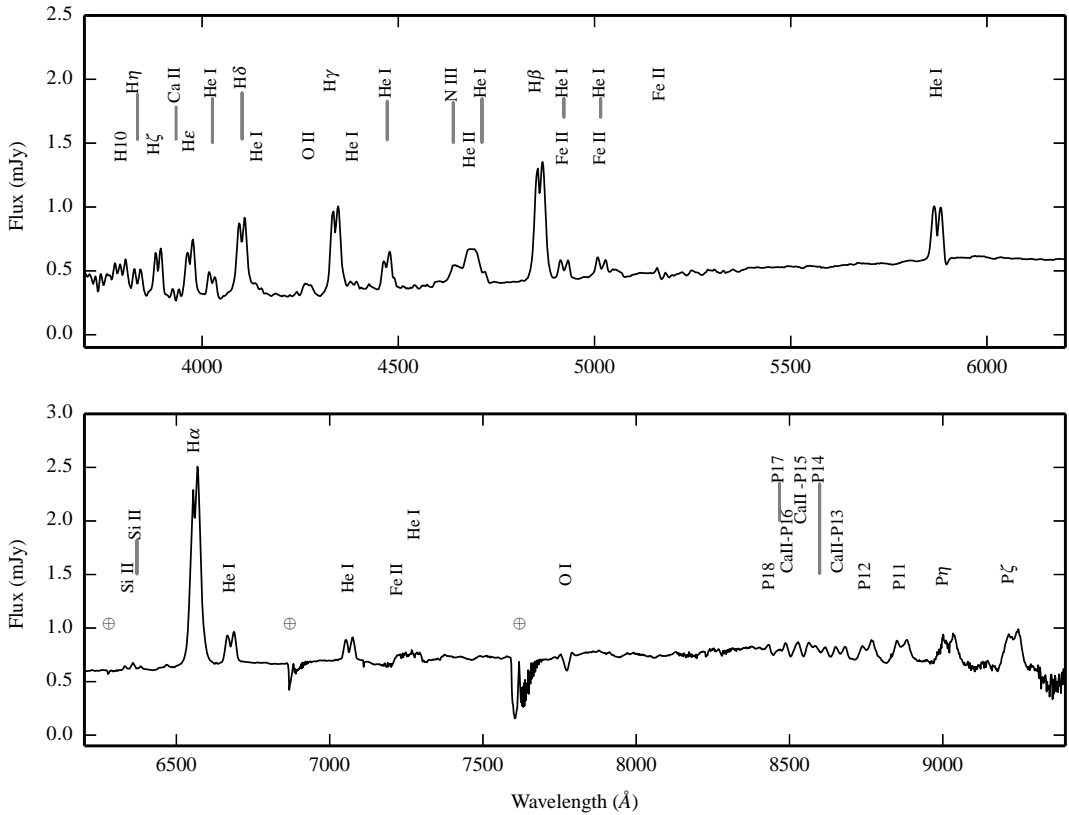}
\caption{Average spectrum of V2051 Oph.  Each of the 43 spectra is given equal weighting and no correction for orbital motion has been made.}
\label{average_voph}
\end{figure*}

The average spectra of our targets are shown in Figures  \ref{average_ccscl} and \ref{average_voph}. We averaged all flux-calibrated exposures, giving equal weighting to each spectrum. Key spectral features are labelled. Telluric lines are not subtracted, but marked with the symbol $\oplus$.  No correction for orbital motion has been made in constructing these average spectra. They were derived from 26 exposures spanning $1.1$ orbits for CC Scl and 43 exposures covering $2.3$ orbits in the case of V2051 Oph.  All lines are in emission and double-peaked.  The spectra are dominated by the Balmer lines that are clearly resolved up to H10 as well as broader lines from the Paschen series from P$\zeta$ to P16. They also show numerous He~I and He~II lines in emission. Some Fe~II and Ca~II lines, particularly the Ca~II triplet $8498.02$~\AA, $8542.09$~\AA\ and $8662.14$~\AA\, are also present in emission.
All Balmer and He lines are asymmetric, showing a stronger blue peak in the spectrum of CC Scl. In the case of V2051 Oph, the red peak is stronger than the blue, differing from previously published spectra that show a stronger blue peak (\citealt{2008A&A...487..611P}, \citealt{2001MNRAS.323..484S} and \citealt{1986A&A...154..197W}).  No orbital sampling effects are believed to be causing this behaviour, since we observed more than one full orbit. 
%
% There is no signal of Mg II 4481\AA.  This line originates in the atmosphere of the WD and it's expected to be un-blended with He~I 4471\AA\ in low inclination and low mass accretion rate systems. %[DS: would you expect to resolve it here? lines look broad]
%
In comparison with previous  published spectra of CC Scl (in quiescence), (e.g. Chen et al. 2001; Tappert et al. 2004), our higher resolution spectra undoubtedly show the presence of He~II $4685.75$~\AA. Furthermore, even though there is no evidence of the   C~III-N~III blend in its blue wing, as reported during outburst by \cite{1997MNRAS.292..397H} there is a sign of He~I $4713.20$~\AA\ in its red wing, previously unresolved.
We also see weak He~II 5411.55~\AA\  and a narrow absorption feature consistent with $4920 $~\AA\, potentially Fe~I is blended with He I.
We constructed light curves from our spectra, but no  eclipse features were detected.   Based on the short eclipse time suggested by Kato's light curve, our time resolution is of the order of half of the total eclipse duration. By covering only one orbital period, and adding the fact that the eclipse is not always visible, it is not surprising that we could not resolve it.

\subsection{Radial velocities analysis}
\label{rvelanal}
\begin{figure}
\centering
 \includegraphics[width=1\linewidth, trim= 0cm 1.6cm 0.6cm 1.2cm, clip=true]{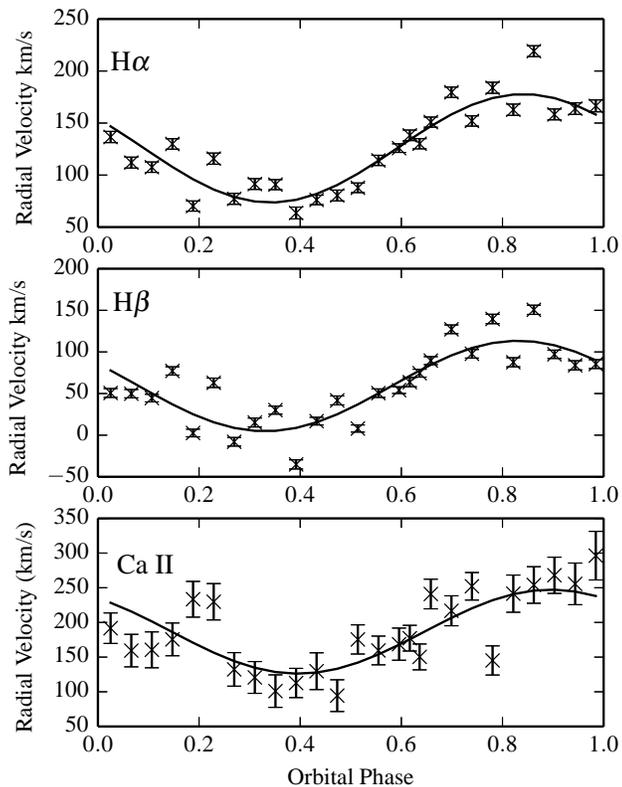}
\caption{H$\alpha$, H$\beta$ and Ca~II radial velocity curves for a Gaussian separation of $2500$~km/s, folded on the orbital period for CC Scl.}
\label{radial_vel}
\end{figure}

\begin{figure}
\centering
 \includegraphics[width=1\linewidth, trim= 0cm 1.6cm 0.6cm 1.2cm, clip=true]{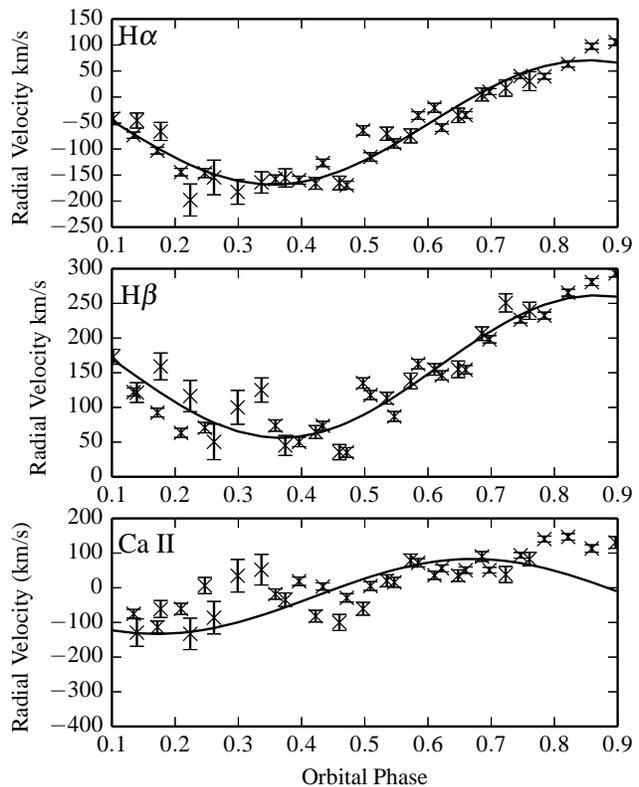}
\caption{H$\alpha$, H$\beta$ and Ca~II radial velocity curves folded on the orbital period, for a Gaussian separation of 3200 km/s  for the Balmer lines and 1800 km/s for Ca~II of V2051 Oph.}
\label{radial_vel2}
\end{figure}

We initially performed the commonly-used double Gaussian method introduced by \cite{1980ApJ...238..946S}, to obtain the radial velocities of several lines. The measured radial velocities were then fitted  with a sinusoidal curve of the form:\\
\begin{center}$v_r(\Phi)=\gamma-K \sin(2\pi(\Phi-\Phi_0))$\end{center}
\noindent  where $\gamma$ is the systemic velocity, $K$ is the radial velocity semi-amplitude of the emission line, $\Phi$ is the orbital phase and$\Phi_0$ is the phase shift.  We let $K$, $\gamma$ and $\Phi_0$  vary. We employed a Gaussian FWHM of $500$~km/s and considered Gaussian separations of $a=500-4000$~km/s in steps of $50$~km/s. This allowed us to construct diagnostic diagrams \citep*{1986ApJ...308..765S} for each line. We compare the two strongest Balmer lines with Ca~II $8662.14$~\AA\ in order to assess how stable the inferred parameters are against the choice of line. The results can be found in Table \ref{tab:tabla_dd}.  The adopted values are the average of all fits values within the flat plateau of $a$, using the full range of the parameters over that range to derive an uncertainty estimate. 
Some representative fits for the reported lines are shown in Figure \ref{radial_vel}. The radial velocity curve of Ca~II has a similar behaviour to H$\alpha$ and H$\beta$ in the case of CC Scl.    These show similar amplitude and phase shifts but significantly different values for the systemic velocity $\gamma$.  It is noticeable in the fits for V2051 Oph (Figure \ref{radial_vel2}), that the Ca~II line results of the diagnostic diagram are particularly poor, suggesting the method is not reliable for this line in the case of V2051 Oph. This is at least partially due to the fact that Ca~II is weaker in V2051 Oph compared to CC Scl. 
% 
% We believe that the large $\gamma$ value derived from the Ca~II line is due to Paschen 13 contamination.  We based this observation on the fact that the $\gamma$ value is around $100$~km/s larger than the average $\gamma$ of the Balmer lines, an alteration consistent with P13 contamination. The value derived for the systemic velocity of V2051 Oph with H$\alpha$ leads us to believe that the  strong bright spot signal on H$\alpha$, as seen in the first panel of Figure \ref{ccscl_dopmaps} misconducted the results towards very high values of $\gamma$.  
%
The discrepant results for $\gamma$ depending on the line are not rare among CVs.  In the case of V2051 Oph, see for example \cite{1986A&A...154..197W}, \cite{2008A&A...487..611P} and references therein.  This behaviour is normally associated with blending of nearby lines or asymmetries between the red and blue wings, both of which are applicable to our targets. 
The asymmetry in the average line profiles, as well as the fact that some of our RV fits are poor due to a distorted RV curve, suggest that these parameters may be skewed by an asymmetry in the disc emission, such as caused by a bright spot. This is a notorious issue with the double-Gaussian method, even when a diagnostic diagram is considered. 
% %
% Since phase delays relative to the expected motion of the primary star can be observed in radial velocity curves of CVs (probably due to residual hot spot contamination), we will derive the absolute phase zero by other means later in this paper (section \ref{secondary}).
% %

\begin{table*}
\centering

\begin{tabular}{l c c c c c c}
\hline
\hline
 & &   CC Scl & & &  V2051 Oph  & \\
 Line      &  $\gamma$~km/s  &   $K_1$~km/s &  $\Phi_0$    &   $\gamma$~km/s  &  $K_1$~km/s &   $\Phi_0$  \\ 
\hline
 H$\alpha$ &  $117\pm20 $   &  $53\pm5   $ &$0.1\pm0.04$  &  $102\pm100 $   &  $120\pm16   $ &$0.1\pm0.02$ \\
 H$\beta$   & $ 55\pm5$      &   $59\pm7  $   &$0.1\pm0.03$ & $ -40\pm40$      &     $114\pm20  $   &$0.0007\pm0.02$ \\

 CaII   &     $193\pm19$ &   $52\pm11  $  &$0.1\pm0.03$      &         $5\pm22$ &     $95\pm28  $  &$-0.01\pm0.05$     \\
\hline
\end{tabular}
\caption{Orbital parameters derived from the diagnostic diagrams of CC Scl and V2051 Oph. The reported errors are the dispersion between the smallest and largest values for each parameter of the averaged plateau range.  The results for Ca~II in the case of V2051 Oph are not reliable, see the bottom panel of Figure \ref{radial_vel2}.}
\label{tab:tabla_dd}
\end{table*}

\subsection{Doppler tomography}
\label{ccscl_dt}

\begin{figure*}
\centering
 \includegraphics*[width=\textwidth, trim= 0cm 0.5cm 1.5cm 0cm, clip=true]{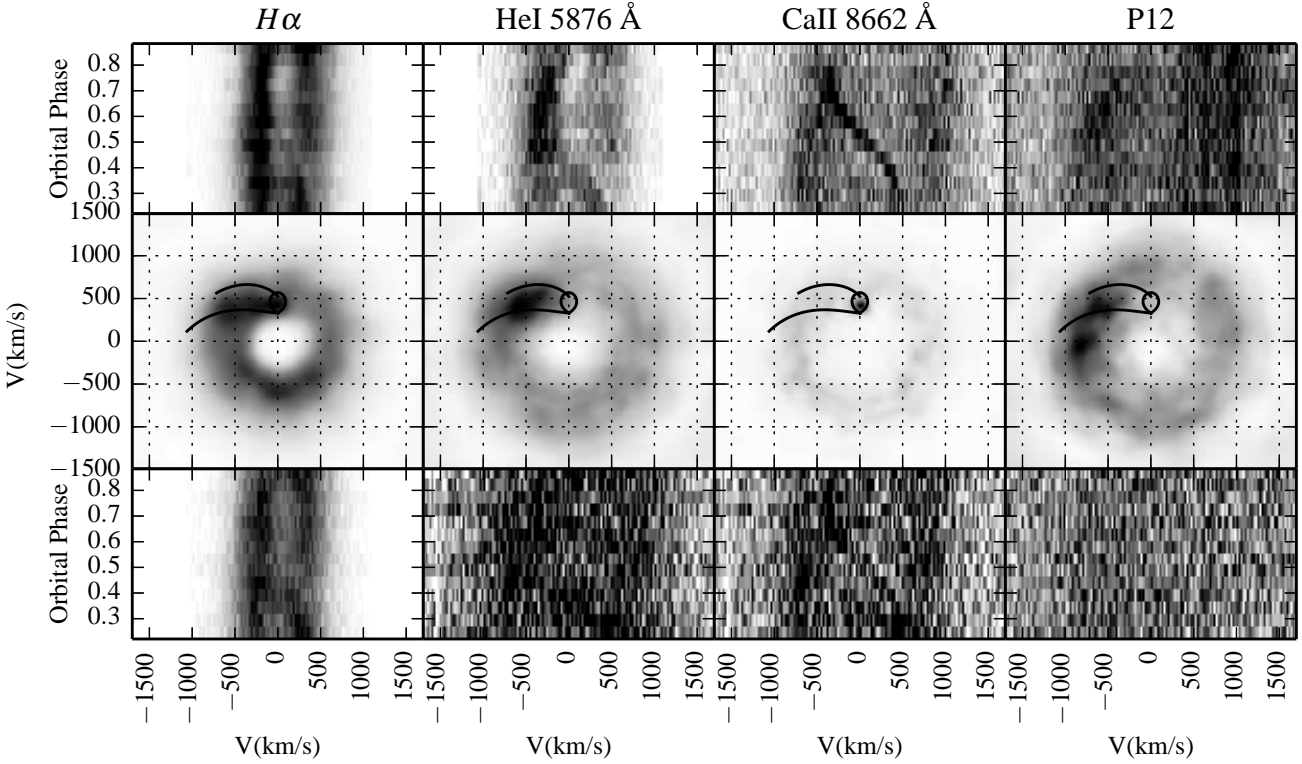}
\caption{ Doppler maps of CC Scl. From top to bottom: Original data, Doppler map and reconstructed data. From left to right: H$\alpha$, HeI $5876$~\AA, Ca~II $8662$~\AA\ and Paschen 12.  The $x$ and $y$ axes are $V_x$ and $V_y$ respectively.  The Roche lobes model for our best orbital solution (Section \ref{k_cor}) and gas stream are plotted with a solid line over the Doppler maps.}
\label{ccscl_dopmaps}
\end{figure*}

\begin{figure*}
\centering
 \includegraphics*[width=\textwidth, trim= 0cm 0.5cm 1.5cm 0cm, clip=true]{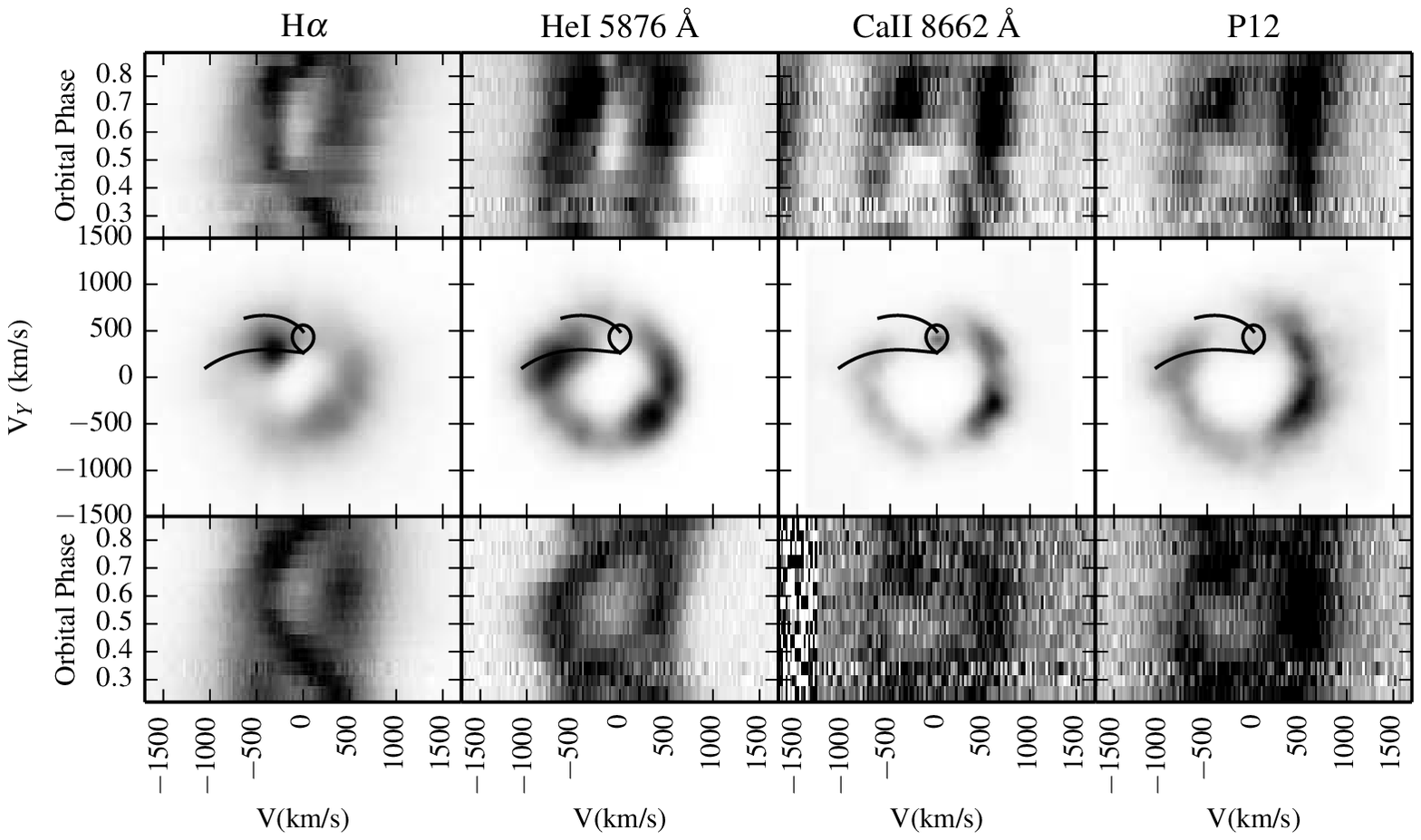}
\caption{Doppler maps of V2051 Oph. From top to bottom: Original data, Doppler map and reconstructed data. From left to right: H$\alpha$, HeI $5876$~\AA, Ca~II $8662$~\AA\ and Paschen 12.  The $x$ and $y$ axes are $V_x$ and $V_y$ respectively.  The Roche lobes model for the best orbital solution (Section \ref{k_cor}) and gas stream are plotted with a solid line over the Doppler maps.  The Ca~II map was led toward low values of $\chi^2$ to enhance the secondary detection.}
\label{voph_dopmaps}
\end{figure*}

We used the indirect imaging technique, Doppler tomography \citep{1988MNRAS.235..269M}, to resolve the origin and kinematics of the various emission components.  Doppler tomography translates phase-resolved emission line profiles into a distribution of line emission in two dimensions of velocity space. Apart from resolving the accretion discs and directly revealing any asymmetries or bright spots within them, it can also reveal emission features from other components, such as the secondary star.
%
% We used Tom Marsh's software \emph{molly} and \emph{doppler} to generate all the maps in this work.
%
The eclipse phases were removed before calculating these Doppler maps. Due to the extensive wavelength coverage, a large number of lines were accessible for tomography, but we only show the most representative lines for each system. All the individual spectra were normalised to the continuum using low-order polynomial fits and the maps were allowed to reach values of $\chi^2 < 2$. 
In Figures \ref{ccscl_dopmaps} and \ref{voph_dopmaps}, we compare the observed data in trailed spectrogram format, with the reconstructed Doppler maps as well as the predicted data in order to visually assess the goodness of fit.
All trailed spectra show clear evidence for a strong disk component, responsible for the double-peaked line profiles and clearly visible in all the reconstructed maps.  In addition to the accretion disc, most lines show enhanced emission in the fourth quadrant ($-V_x,+V_y$), identified as the bright spot.  The bright spot seems to be consistent with the trajectory of the gas stream if we consider the expected binary parameters.  In addition to disc and bright spot contributions, the maps of V2051 Oph show additional regions of enhanced emission in the disc. In many lines, the right side of the disc is even brighter than the bright spot region.  This enhanced emission has been reported before for this system while coming out of superoutburst by \cite{2008A&A...487..611P} and it was believed to be the equivalent of the superhump light source. It is noticeable that this emission is located in the same place as it was when previously reported by \cite{2008A&A...487..611P}.  
We reconstructed Doppler maps for the six stronger He~I lines.  They all showed the same features and show the brightest line, HeI $5876$~\AA\ for both systems.

%The Roche lobes model for the best orbital solution (calculated later in this paper) and gas stream are plotted with a solid line over the Doppler maps.  [do this in the Figure caption]
A sharp S-wave moving in between the disc peaks is noticeable in the trailed spectra of Ca~II .  This feature is from the secondary star, as is clearly visible in the Doppler map of Ca~II of CC Scl (third panel of Figure \ref{ccscl_dopmaps}).  This Doppler map was pushed towards low $\chi^2$ values in order to enhance the secondary star feature. Traces of the secondary emission are also visible in the maps of the Balmer lines from H$\alpha$ to H$\delta$, at velocities $(V_x,V_y)=(0,410)$~km/s in the case of CC Scl.  The emission is strongest and sharpest in the Ca~II lines, as shown in the third panel of Figure \ref{ccscl_dopmaps}.  The disadvantage with Ca~II triplet is the close proximity to the Paschen lines P13, P15 and P16, respectively. This makes the Ca~II profiles blend with the Paschen lines as mentioned before.  We have measured the contribution of each Paschen line to the respective Ca~II line by estimating the trend from Paschen 10 to 12. The contribution of P13 to the Ca~II maps is 50\% of the disc flux. After performing some tests, we found that the best way of ensuring that the blending did not distort or limit our ability to reconstruct Ca~II maps was to calculate the ratio between the blends and reconstruct simultaneous Doppler maps of both Ca~II and Paschen.  The map shown in the third panel of Figure \ref{ccscl_dopmaps} is the corresponding Ca~II de-blended  map. Since P12 is not blended with any neighbouring lines, we show its Doppler map as representative of the Paschen lines. The map of P13 shares the same main features of P12 but includes an enhanced flux in the first quadrant that we identified as residual contribution of the secondary emission from Ca~II. The absence of a secondary feature in the P12 map confirms that this component is solely present in Ca~II.
%

%As discussed before, the Ca~II Doppler map is the best tracer of the secondary's position and will be used in further analysis.  

\subsubsection{Systemic velocity $\gamma$}
\label{dopmap_gamma}

\begin{figure}
\centering
 \includegraphics[width=1\linewidth]{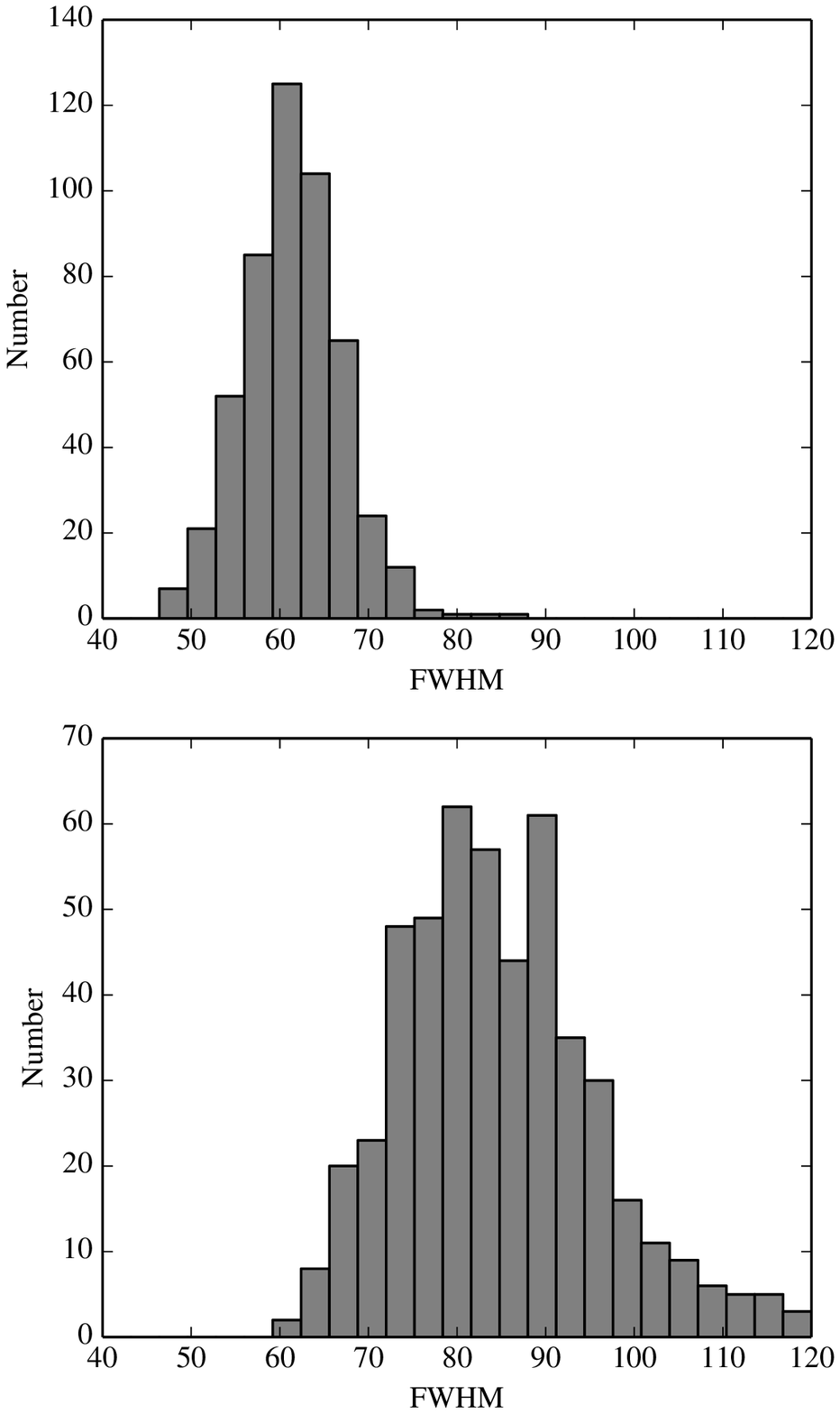}
\caption{Histograms of the measurements of FWHM of the secondary star of V2051 Oph for two values of $\gamma=0$ and $\gamma=50$~km/s }
\label{gamma}
\end{figure}

\begin{figure}
\centering
 \includegraphics[width=1\linewidth]{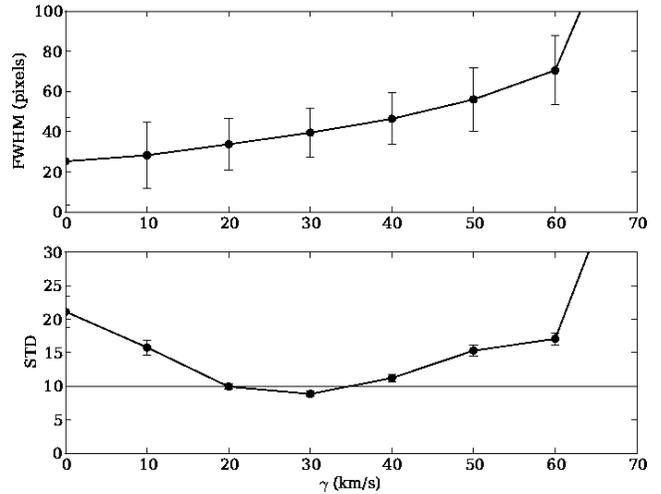}
\caption{FWHM of the secondary star and STD versus value of systemic velocity $\gamma$ for CC Scl.  The error bars are sometime smaller than the symbol.}
\label{ccscl_gamma}
\end{figure}

\begin{figure}
\centering
 \includegraphics[width=1\linewidth]{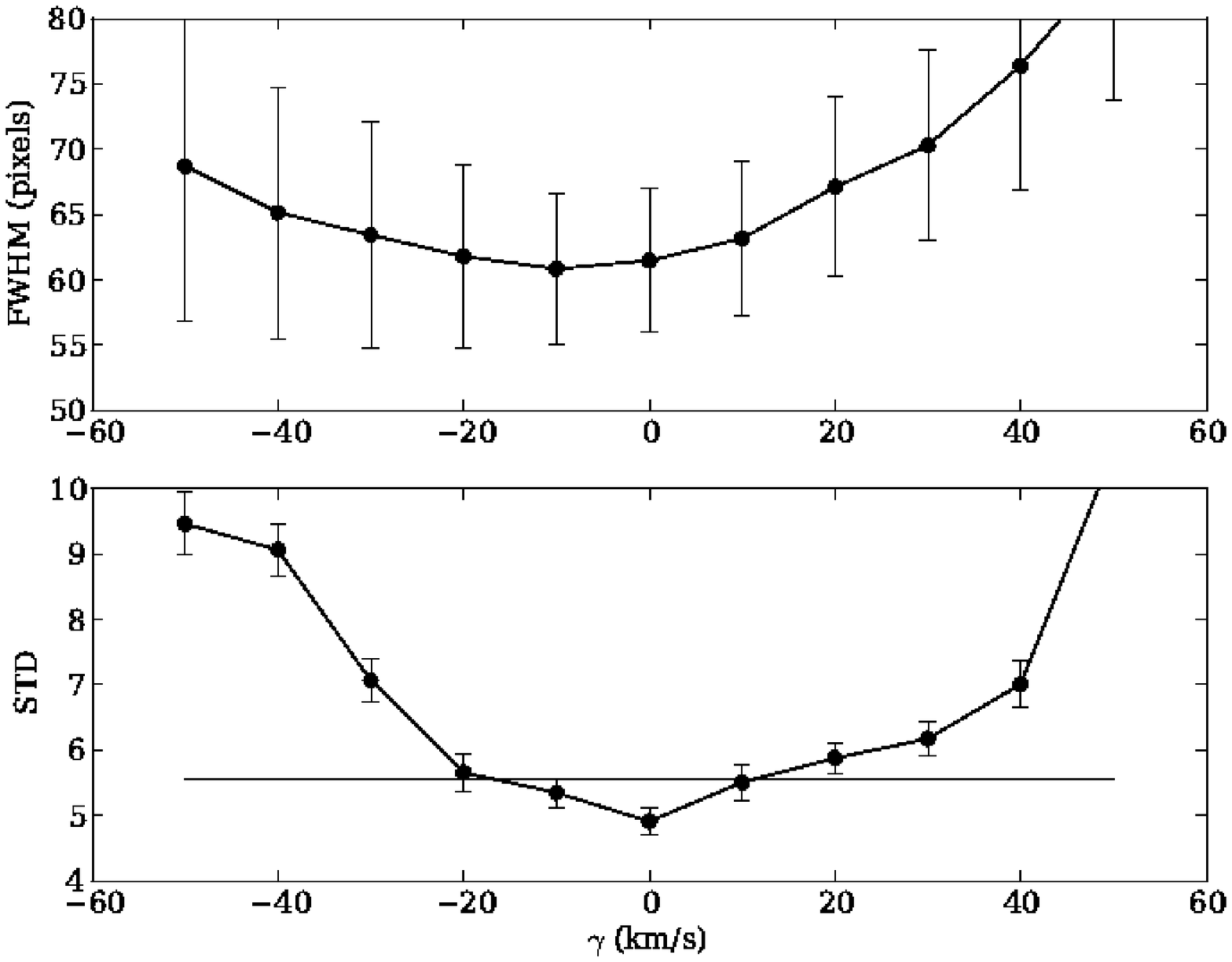}
\caption{FWHM of the secondary star and STD versus value of systemic velocity $\gamma$ for V2051 Oph.}
\label{voph_gamma}
\end{figure}

As the measurements of $\gamma$ from the diagnostic diagram didn't yield consistent results, we determined the systemic velocity using the Doppler maps of Ca~II $8662.14$~\AA. Our method relies on the dependence of the maps on the assumed systemic velocity. It is the only input parameter to Doppler mapping, though in the case of maximum-entropy regularised mapping as employed here, the user also chooses the target $\chi^2$. An incorrect $\gamma$ leads to poorer reconstructions and if sharp S-waves are present will lead to a noticeable blurring of the reconstructed emission source of such an S-wave. Conversely, the correct value of $\gamma$ will generate maps with optimal sharpness.  The detection of a sharp S-wave from the donor offers the opportunity to  study the reconstructions as a function of the assumed $\gamma$.

We generated a series of maps with systemic velocities varying from $-200$ to $200$~km/s in steps of $10$~km/s. The emission spot of the secondary was fitted with a 2D Gaussian in order to determine the width of the spot.
We choose Ca~II over the Balmer lines since it contains a sharp S-wave from the donor that should map to a sharp spot in the maps. This initial search was used to determine an approximate estimate for the systemic velocity, $\gamma$. We then performed a search for $\gamma$ spanning $\pm50$ km/s around this estimate in steps of $2$~km/s. Furthermore, instead of reconstructing one map per assumed gamma, we bootstrapped our data to generate 500 independent data-sets \citep{efronboot}, and reconstructed Doppler maps for these data while aiming for the same $\chi^2$ as the original map. In each map, the emission spot is fitted such that we can study the distribution of spot parameters. 
For each assumed $\gamma$, we construct a histogram of the FWHM distribution of the 500 spot measurements. 
Figure \ref{gamma} show two examples of these histograms in the left panels comparing $\gamma=0$~km/s and 50~km/s.  

Since these histograms are well represented with a Gaussian distribution, we fit a Gaussian to these and determine the mean and $\sigma$ of these distributions. Figure \ref{gamma} shows that the average value of the FWHM is smaller for the $\gamma=0$~km/s case, also the distribution with the smaller value of FWHM has a narrower width. This is expected if $\gamma=0$~km/s is closer to the true value than $\gamma=50$~km/s, as the latter would lead to distorted and broader emission spots in the reconstructions. Thus when considering these two parameters as a function of assumed $\gamma$, we would expect to be able to pick the preferred value for the systemic velocity. In Figure \ref{ccscl_gamma} and \ref{voph_gamma} we show how these parameters evolve as a function of the systemic velocity for our two targets.  The case of V2051~Oph is straightforward as both the FWHM as well as $\sigma$ minimise for $\gamma=0$~km/s. For CC~Scl, a clear minimum is found in the sigma distribution and thus we adopt the minimal value of sigma  for our best gamma in both cases.
To estimate an uncertainty on this optimal $\gamma$, we consider the error derived on the standard deviation of the distributions and consider all the values of the systemic velocity whose sigma values are less than three times the standard deviation away from the optimal $\gamma$ (horizontal line cutting the lower panel of Figures \ref{ccscl_gamma} and \ref{voph_gamma}).  The reported error is then the value of $\gamma$ where the horizontal line cuts the STD.
Using this recipe, we find a value of $\gamma=30_{-22}^{+33}$~km/s for CC Scl and $\gamma=0_{-17}^{+12}$~km/s for V2051 Oph. These are the values that we adopted to generate all the Doppler maps shown in this paper.

\subsubsection{Secondary star and phase zero}
\label{secondary}

As noted above, emission from the donor stars is easily identified in the Doppler tomograms and best seen in the Ca~II lines (Figure \ref{ccscl_dopmaps} and \ref{voph_dopmaps}).  This was first demonstrated for GW~Lib in \cite{2010MNRAS.401.1857V}, and in both of our targets the donor star emission is both stronger and sharper in the Ca~II triplet compared to other lines. 
The detection of the donor star in emission offers the first direct constraint on its orbital velocity for our targets. Deriving this velocity from the Doppler maps has the advantage that such a component can be isolated in velocity-space easily even when the S-wave from the donor is too weak to permit direct fits to the data. This is because Doppler tomography fits to all the data simultaneously, in the co-rotating frame of the binary, ideally suited for faint donor star signals.  Another advantage is the ability to model blended lines, as is the case for Ca~II.
We fitted 2-dimensional Gaussian profiles to all our donor emission spots in order to determine its  $V_x, V_y$ position in Doppler coordinates and thus determine the corresponding radial velocity semi-amplitude $K_{em}$. 
The detected emission give us $K_{em}=402\pm 6$~km/s for CC Scl and  $K_{em}=389\pm 11$~km/s in the case of V2051 Oph. The errors were derived using a bootstrap method as described in the previous section. The distribution of $K_{em}$ values across our 500 bootstraps was Gaussian.
% %
% As the secondary emission is likely powered by irradiation, $K_{em}$ does not track the centre of mass of the secondary and can be significantly smaller if the emission is concentrated towards the L1 point. To estimate the real velocity of the donor ($K_2$) we will construct irradiation models and correct for the emission component in the next section.

The detection of the donor also allows us to define the absolute phases of the systems. With the correct ephemeris, the secondary star is projected at Doppler coordinates $(V_x,V_y)=(0+K_2)$. A phase error amounts to a rotation in Doppler maps, and thus we determined the rotation needed to align the secondary in order to define phase zero. Here we again made use of our 500 bootstraps to be able to determine this phase and its error robustly. We find:

\noindent HJD= $2455384.9869\pm0.0003$
 for CC Scl, and\\
HJD= $ 2455384.0124\pm0.004$
for V2051 Oph.\\

\noindent  All the maps in this paper are generated with the above reported phase reference. Since there is a well established eclipse ephemeris for V2051 Oph, we compared our values with the ephemeris of \cite{1998MNRAS.300..233B} and find that they agree within $3\sigma$.  This confirms that the emission spot we used to set phase zero does indeed track the donor star. Recently, \citep{vsnet} reported an ephemeris for CC Scl on the basis of eclipse features seen in outburst light-curves, and again we find good agreement (within $3\sigma$). This does suggest that CC Scl is a high-inclination system even though we do not see obvious evidence for an eclipse in our data.

\subsubsection{Primary velocity $K_1$}
\label{k1}

In section \ref{rvelanal}, we applied the standard radial velocity treatment to find the radial velocity of the white dwarf by means of tracing the velocities of the accretion disc.
In Doppler maps, the WD is located at $(V_x,V_y)=(0,-K_1)$. In principle, the disc feature in Doppler maps should also centre on the white dwarf position. For example, a pure and azimuthally symmetric Keplerian disk should map to a ring of emission centred exactly on the WD. Thus a centre of symmetry search of the disk component in Doppler maps should be able to constrain $K_1$. The centre of symmetry technique has been successfully used for Her X-1 \citep{1997MNRAS.292...52S}, Sco X-1 \citep{2002ApJ...568..273S}, WZ Sge \citep{2007ApJ...667..442S} and GW lib \citep{2010MNRAS.401.1857V}.

For an assumed centre, we first determine the symmetric component of the reconstructed map, which is then subtracted from the map. This asymmetric map is then inspected for residuals. The optimal centre of symmetry minimises these residuals. The presence of non-disc components or strong asymmetries can distort such a residual map and thus bias the measurement. However, these can be easily masked or ignored, so that the minimisation focuses on residuals near clean disk regions rather than a minimisation across the whole map. This is a significant advantage over double Gaussian methods where such distortions are difficult to remove, even if indirect evidence for them is obvious in the form of, for example, phase shifts.

Henceforth, we will call this region of Doppler space over which we minimise the residuals our ``box''.  It is important not to over-fit the maps as it would break the disk component up into a large collection of spots. Thus we tune the target $\chi^2$ to ensure a relatively smooth disk structure is reconstructed. 
To test the behaviour of the method depending of the chosen line, box size and box location, we considered different box regions and explored the five strongest Balmer lines and Ca~II 8662\AA. An obvious region to exclude is the quadrant containing the bright-spot, as well as the region near any donor star emission.
Good regions are the bottom left quadrant ($-V_x$,$-V_y$) and the top right quadrant ($+V_x$,$+V_y$).  We also tried  a more extended area including the bottom right quadrant despite seeing slight asymmetries in some lines, and rectangular boxes covering the bottom half and the right half of the map. We also ran the analysis on a box that effectively uses the whole map, to further quantify the impact of the chosen region on the inferred $K_1$.

We searched over a range of possible symmetry centres, changing both the X and Y position of our centre. We first varied the $y$-position over a range of 200~km/s in steps of 10~km/s holding $x$ at zero. At the optimal $y$, we then varied $x$ over a similar range. We then allowed $y$ to vary again and re-optimise for $y$ to deliver the final best centre of symmetry position $(K_x,K_y)$, and thus an estimate for $K_1$ via $K_1=\sqrt{K_X^2+K_y^2}$. This sequence was repeated for all lines and maps under consideration as well as over a range of regions. Although one would expect the $x$-component to be zero if the disk was fully symmetric.  We do find small, but non-zero values for $K_x$, suggesting that some contributions from disk asymmetries remain.

The lines that showed the most stable behaviour were Ca~II 8662\AA\ and H$\beta$. Variations between them for the same box are $\sim20$ km/s, variation between different boxes are $\sim20$ km/s in the half quadrant boxes, and up to $\sim50$ km/s in the rest. The higher Balmer lines showed slightly larger variations, while H$\alpha$ didn't deliver consistent values.  We believe that the lack of consistency of H$\alpha$ could be due to the strength of the bright spot feature in this line.  

With this technique, there is a compromise between the size of the box and the impact of asymmetries. A larger region uses more of the data, but has a higher chance of including asymmetries and non-disk components that impact the apparent centre of symmetry. We found that using a box covering around a half of the Doppler map, trying to avoid the most asymmetric parts of the disc, gives the best results for all the studied lines. 

In the case of CC Scl, Ca~II Doppler maps show a well-defined  accretion disc, permitting a more accurate estimate for $K_1$.   In the case of V2051 Oph, even the Ca~II Doppler map was very asymmetric (Figure \ref{voph_dopmaps}). We re-examined the maps and found that the $H\delta$ map was not as sharp as Ca~II but featured a relatively symmetric disk (see Figure \ref{voph_disc}). After this careful comparison of different lines and regions, we finally settled on using the right half of the Ca~II map for CC Scl and the bottom half of the $H\delta$ map for V2051 Oph. Having established these optimal line/region combinations, we then repeated the analysis using 500 bootstrap data samples and performed a symmetry search for each reconstructed map. The mean value was adopted as our best estimate for $K_1$, using the standard deviation as its error. This delivered $K_1=37\pm14$~km/s for CC Scl and $K_1=97\pm10$~km/s for V2051 Oph. The former is consistent with the results from the diagnostic diagram, while the latter is in between the expected value from the eclipse solution ($K_1=83\pm12$) and that derived from the diagnostic diagram ($K_1=110\pm16$). It is consistent with the eclipse solution given the uncertainties and appears to be a better estimate than we were able to derive from the diagnostic diagram. This puts some confidence in the method and suggests it is a useful tool for constraining $K_1$, with some clear advantages over the traditional double Gaussian approach, including the ability to deal with obvious asymmetries in a straightforward manner and a more robust uncertainty by using bootstrapped reconstructions.

\subsubsection{The binary mass ratio}
\label{k_cor}

Our detection of the donor star in emission requires a correction before it can be used to determine the donor star's orbital velocity ($K_2$).  When combined with our $K_1$ constraint, this delivers the binary mass ratio, an important parameter for establishing the evolutionary state of a CV.
The emission from the donor is clearly phase-dependent in strength, peaking at phases where we look at the side facing its white dwarf companion, consistent with emission powered via irradiation. Since this side of the donor star orbits at lower radial velocities than its centre of mass, a positive correction is required to estimate the true $K_2$; $K_{2} =  K_{em} + \Delta K$.
In the extreme case that the emission is centred on the L1 point of the donor, this deviation would be maximal. The other extreme is no correction at all. Either scenario is very unlikely and thus we have to determine an appropriate $\Delta K$ with $K_{L_1} < K_{em} < K_2$.

%Some long period CVs have reported slingshot prominences present shortly after or during outburst (e.g. IP Peg and SS Cyg: \citealt{steeghs96-1}, BV Cen: \citealt{watsonetal07-1}). In these cases,  the emission is associated with the donor, but enclosed within a magnetic loop extending into the primary Roche lobe. As these loops tend to find equilibrium near the centre of mass, they have a much smaller radial velocity component than $K_2$. More recently, similar emission features have been found in short period CVs in quiescence (e.g. SDSS J003941.06+005427.5: \citealt{southworthetal10-1}). As these features have \emph{very low} radial velocities we can safely assume that the origin of the emission seen in this work lies on the donor surface and is constrained between the inner Lagrangian point and the centre of the donor star such that $K_{L_1} < K_{em} < K_2$. 
%
We used the \emph{lprofile} code from the {\sc lcurve} package that calculates synthetic line profiles for an irradiated secondary (e.g. \citealt{2010MNRAS.402.1824C},  \citealt{2010MNRAS.402.2591P}).  Synthetic phase-resolved data were generated for a range of models and fitted to derive the equivalent $K_{em}$ that one would measure if this was the correct model for the secondary. For a given system, only models that are consistent with the observed $K_{em}$ are viable, and the true $K_2$ is then known from the model. The key input parameters in these models is the assumed binary mass ratio, the inclination and any shielding of the secondary star due to the presence of a vertically extended and optically thick disc. 
Tests with model data confirm that the assumed inclination has little impact, delivering $K_2$ to within a few percent over the full inclination range. Our findings are thus consistent with what was previously reported by \cite{2005ApJ...635..502M} in their extensive study of the K-correction using a different code. We  fixed the inclination of the model to the system's value in the case of V2051 Oph, while for CC Scl (unknown inclination, but shallow eclipses) we fixed the model inclination to $80^{\circ}$.
The correction is obviously strongly dependent on the, to be determined, mass ratio and the unknown accretion disc thickness can be a significant effect due to shielding, but in our case the mass ratio is the dominant parameter. 

\begin{figure}
\centering
 \includegraphics[width=1\linewidth,trim= 0.4cm 0.2cm 1.1cm 1cm, clip=true]{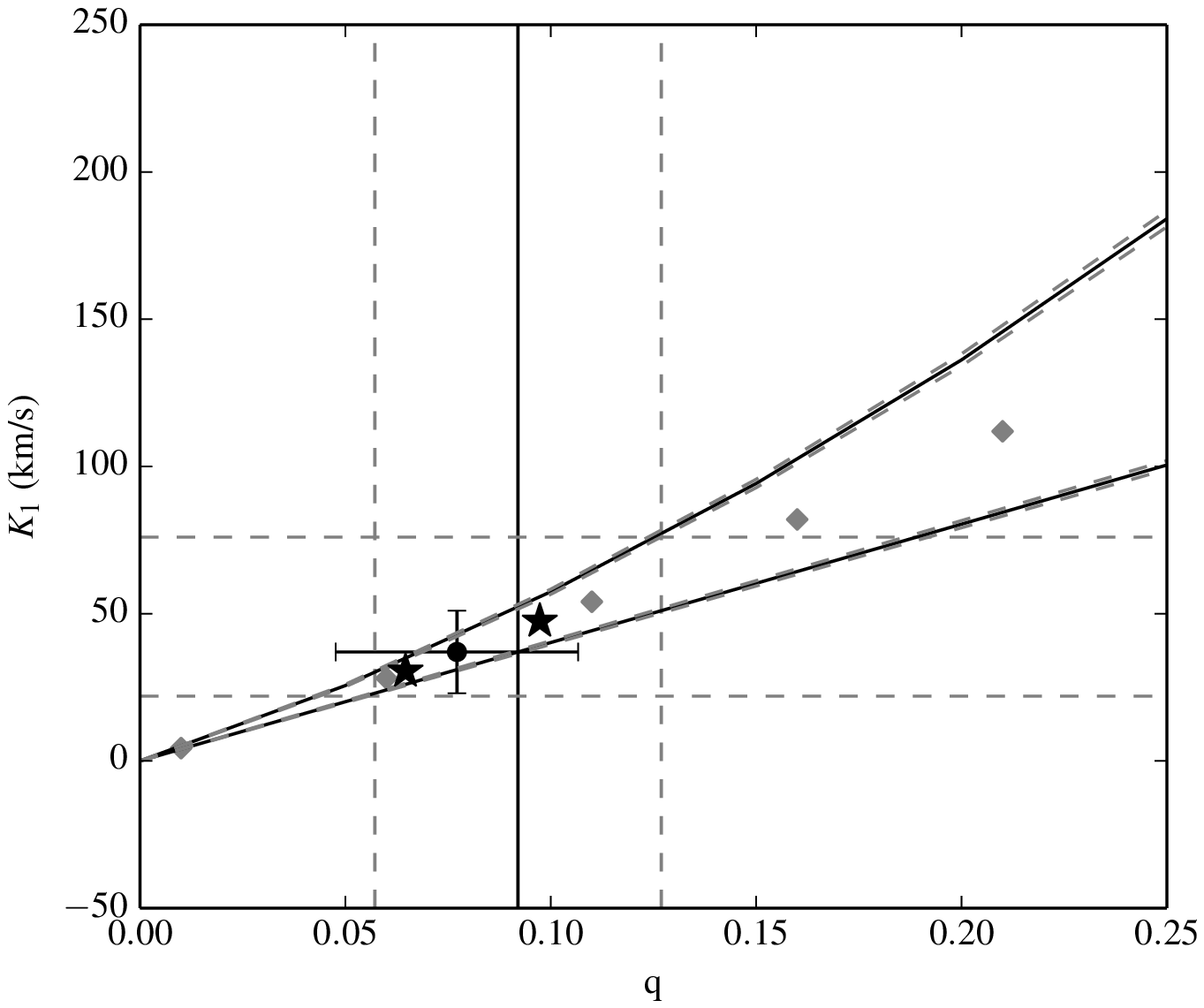}
\caption{$K_1$, q plane for CC Scl. $K_{em}=402\pm6$~km/s, $K_{em}>K_{L_{1}}$ is the top diagonal solid black line with dashed errors.  $K_{em}<K_2$ is the bottom diagonal line with dashed errors. The vertical solid line represents the maximum value of q.  The horizontal dashed grey lines are the highest and lowest estimates of $K_1$ found for four Balmer lines. The grey diamonds are the solutions from our model.  The black circle with error bars represents our best solution for the value of $K_1$ found from the centre of symmetry ($K_1=37\pm14$). The black stars mark the pre- and post-bounce solutions (See section \ref{discussions}).}
\label{ccscl_kcorr}
\end{figure}

\begin{figure}
\centering
 \includegraphics[width=1\linewidth,trim= 0.4cm 0.2cm 1.1cm 1cm, clip=true]{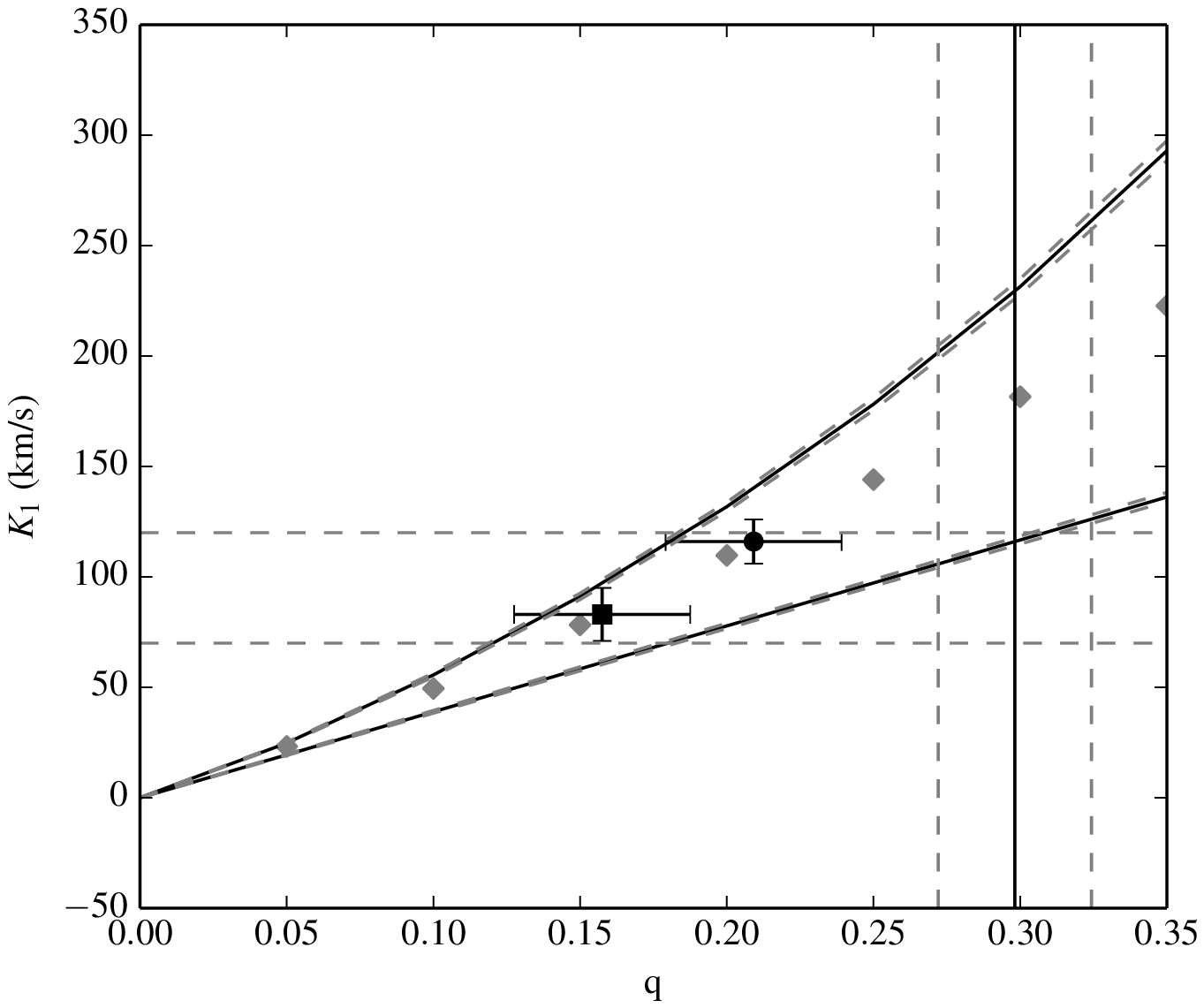}
\caption{$K_1$, q plane for V2051 Oph. $K_{em}=389\pm11$~km/s, $K_{em}>K_{L_{1}}$ is the top diagonal solid black line with dashed errors.  $K_{em}<K_2$ is the bottom diagonal line with dashed errors. The vertical solid line represents the maximum value of q.   The horizontal dashed grey lines are the highest and lowest estimates of $K_1$ found for four Balmer lines. The grey diamonds are the solutions from our model.  The black circle with error bars represents our best solution for the value of $K_1$ found from the centre of symmetry ($K_1=97\pm10$). The black square   with error bars represents the best solution with the value of $K_1$ calculated from the eclipse ($K_1=83\pm12$, see section \ref{back}).}
\label{voph_kcorr}
\end{figure}

We visualise our dynamical constraints in Figures \ref{ccscl_kcorr} and \ref{voph_kcorr} where we show the $K_1$-$q$ plane for CC Scl and V2051 Oph. We represent the constraint  $K_{L_{1}}<K_{em}<K_2$ as diagonal black solid lines with dashed grey error bars. A hard upper limit on the mass ratio is provided by $q_{\mathrm{max}} = \frac{K_{\mathrm{disc}}}{K_{\mathrm{em}}}$ since $K_1 < K_{\mathrm{disc}}$ and $K_2 > K_{\mathrm{em}}$.  This constraint is represented by the black vertical solid line with dashed errors. The horizontal dashed lines indicate  the full range of $K_1$ values (see section \ref{k1}).
Our irradiation models allow us to make a good estimate for the K-correction as a function of mass ratio, as illustrated by the grey  grid diamonds in Figures \ref{ccscl_kcorr} and \ref{voph_kcorr}. We consider the preferred solution to be the model that reproduces the observed $K_{\mathrm{em}}$, while being consistent with the other constraints. It is the error on $K_{\mathrm{em}}$ and $K_{\mathrm{1}}$ that then defines the allowed parameter range for the binary mass ratio.

In the case of CC Scl,  we found that for a $K_{em}=402\pm6$~km/s and $K_1=37\pm14$~km/s our best solution is $q=0.08\pm0.03$.  These values are in good agreement with the findings of \cite{2001MNRAS.325...89C} (Section \ref{back}).

V2051 Oph has a full binary solution in the literature since it is eclipsing and can thus be used as a benchmark for our method. Given $K_{em}=389\pm11$~km/s and $K_1=97\pm10$~km/s we find $q=0.18\pm0.05 $. Had we adopted the $K_1$ implied  by the eclipse solution, we would obtain $q=0.16\pm 0.03$. Thus both our $K_1$ and $q$ are consistent with the reported eclipse solution, giving us confidence in our methodology. While in the case of V2051 Oph, the eclipse solution is to be preferred, only a few eclipsing CVs are known at short periods, while our method has the promise to deliver dynamical mass ratios for a much larger sample of systems.

\section{Discussion and Conclusions}
\label{discussions}

Our analysis has demonstrated that the Ca~II triplet lines offer significant advantages over the more commonly used Balmer and helium lines. Despite being weaker than these lines, the triplet offers a superior view of the accretion disk and appears to be very sensitive to irradiated donor stars. 
Computational power now permits the use of bootstrap Monte-Carlo methods in Doppler tomography whereby a large number of maximum-entropy regularised reconstructions can be performed in a manageable time. As each bootstrap reconstruction is completely independent, it can also very easily be parallelized across multiple cores or cluster nodes. The bootstrap samples allow for a more robust evaluation of the significance of features in Doppler maps and we show how they can be used to perform quantitative analysis on Doppler maps including errors estimation. 
For example, the determination of the radial velocity of the donor star emission that can be seen in maps can be robustly determined by fitting a 2D Gaussian to each reconstruction and then considering the fit parameter distributions across the bootstrap samples in order the derive the optimal velocity and its error.
We also developed an (automated) algorithm that optimises the systemic velocity of the binary $\gamma$, one of the few input parameters to Doppler tomography. Sharp emission sources, such as the donor star, can be used to test for the optimal value of $\gamma$ for which the sharpest and most stable spot reconstruction is expected (see also Casares et al. 2003). 
Finally, we showed how a centre of symmetry search focusing on the disk component in Doppler maps can provide robust constraints on the white dwarf radial velocity ($K_1$). It offers several advantages over the traditional double-Gaussian method given its ability to mask and ignore large asymmetries that typically distort such measurements. It also profits from the bootstrap technique in terms of assigning an uncertainty on the derived $K_1$. 

When these techniques are combined, robust constraints on the binary mass ratio can be derived. However, we must bear in mind that both our tracer of the donor through its emission as well as the tracer of the white dwarf using the disk are not direct measurements of their centre of mass motion. In the case of the donor, we demonstrate that an irradiation model can provide good estimates for the expected K-correction that needs to be added to the observed emission velocity in order to estimate $K_2$. We find that the more difficult parameter is $K_1$ through disk emission. Although we believe our centre of symmetry method to be better at handling disc asymmetries and distortions compared to the traditional double-Gaussian methods, these can never be fully excluded and thus our uncertainties on the binary mass ratios are set by our knowledge of $K_1$. We do find that all of these methods profit from using the Ca~II triplet lines which offer sharp donor star features as well as sharp disc features.

\subsection{CC Scl}

Our mass ratio for CC Scl (q=$ 0.08\pm0.03$) is in agreement with the values  of \cite{2001MNRAS.325...89C}. Using $M_2$ derived by the $M_2-P_{orb}$ empirical relation derived by \cite{Warner1995}, \citeauthor{2001MNRAS.325...89C} calculated a WD mass of $M_1>1.1M_{\odot}$.  If we instead use the more current empirical CVs donor sequence proposed by \cite{2011ApJS..194...28K}, given a $M_1=0.75$, we obtain two values for the mass ratio $q_{pre}=0.097$ and $q_{post}=0.065$.    These solutions correspond to the pre-bounce and post-bounce solutino at the obital period of CC Scl (derived from \citealt{2011ApJS..194...28K}, table 4). Both values are in agreement within error bars with our derived mass ratio which sits in between. We plotted these solutions in Figure \ref{ccscl_kcorr} as black stars. The pre-bounce solution is located over the maximum limit of $q$ and predicts a value of $K_1$ higher than our calculated value.  Given the tendency of our method to overestimate $K_1$, we thus favour the post-bounce solution. According  \cite{2011ApJS..194...28K}, the pre-bounce solution implies an M7 type and the post-bounce a  T type donor star.  Both presumed donor spectral types should be visible in the L band. To further clarify the donor's nature, we suggest infrared time-resolved spectroscopy of the target to attempt a direct measurement of the velocity of the donor.  CC Scl is an interesting target and future observations could confirm its potential post-bounce nature.    

\subsection{V2051 Oph}

\begin{figure}
\centering
 \includegraphics[width=1\linewidth]{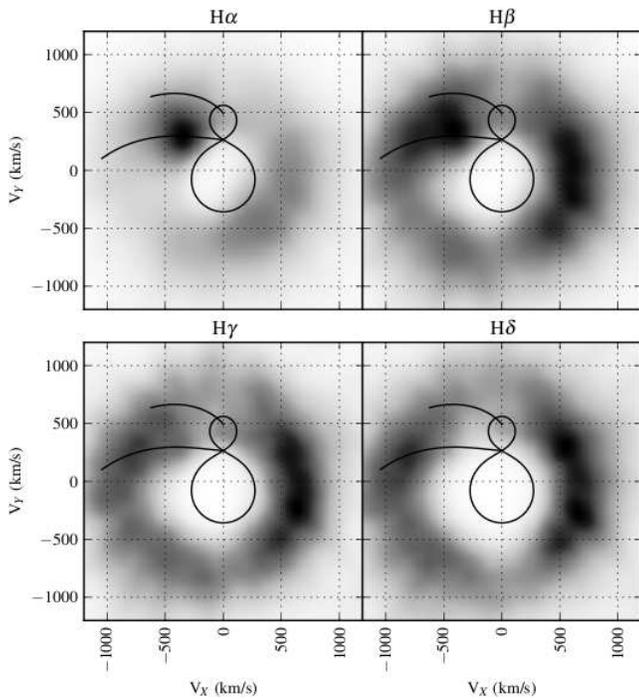}
\caption{Doppler maps of the most prominent Balmer lines of V2051 Oph}
\label{voph_disc}
\end{figure}

As can be seen in Figure \ref{voph_dopmaps} in section \ref{ccscl_dt}, there is  an enhanced emission area in the second quadrant ($+V_x,-V_y$) of the Doppler maps of V2051 Oph.  This feature is not consistent with the bright spot and has been noticed in previous Doppler maps by \cite{2008A&A...487..611P} in the Balmer lines and HeI.  Nevertheless, the two sets of Doppler maps show very different features, as the average spectra had highlighted (Section \ref{spectrum}). The maps from \cite{2008A&A...487..611P},  showed an increase in the bright spot strength when moving to shorter wavelengths and as the bright spot strength decreased, the second quadrant feature increased.  In our case, the reversed relation between the strength of the bright spot and the second quadrant emission is maintained, however, the intensity of the bright spot decreases when moving towards shorter wavelengths in both Balmer lines and HeI (Figure \ref{voph_disc}).  This fact, combined with the discrepancies between the values of $\gamma$ of 
different epochs, makes us believe that the system might be experiencing precession.  Precession periods are typically $\sim 2 d$, leading to variations on timescales longer than the orbital period.  This precession would account for the different values of $\gamma$ found with the double Gaussian method  by \cite{1986A&A...154..197W}, using 2.1 orbits of the system, \cite{2008A&A...487..611P}, using 2.7 orbits and this paper, with 2.3 orbits of the system, indicating that considerably more than two orbital cycles should be necessary to get a reliable value for $\gamma$ with the double Gaussian technique.  We believe that at the time that our data were taken, there was a difference of half a precession cycle with respect to the data of Papadaki et al., explaining for the contrasting lines behaviour.  The value of $\gamma$ derived by our method in Section \ref{dopmap_gamma} is independent of the disc and hence not affected by its variation.

\cite{2008A&A...487..611P} report that their Doppler maps of the Fe II triplet do not show either bright spot or the second quadrant emission mentioned above.  They explain this featureless disc by proposing the existence of a gas region above it where the triplet might originate via fluorescence.   We made Doppler maps of Fe II $5169$~\AA, the only line of the triplet that is not blended with HeI.  This Doppler map, displayed a disc with no signs of the second quadrant emission, but with a bright spot.  The presence of the bright spot in our maps  contradicts this theory of the origin of the Fe II triplet.

As seen in Figure \ref{voph_kcorr}, where the highly asymmetric disc of V2051 Oph led us towards a value of $K_1$ slightly higher than the eclipse solution (black square in Figure \ref{voph_kcorr}), the value of the mass ratio obtained was $q=0.18\pm0.05$, in agreement with the eclipse solution,$q=0.19\pm0.03$, building up confidence in the performance of our K-correction method. 

\vspace{1cm}

 In summary, Doppler map  based methods can provide strong constraints on the orbital parameters of short period CVs.  We compared these methods against classic double Gaussian methods finding advantages from the former over the classic ones.  Our Doppler tomography based methods gives good estimates for the values of $\gamma$, $K_1$ and $K_2$, but our optimal value of $q$ will still depend upon the accuracy of $K_1$.

\section*{Acknowledgements}

The authors acknowledge the use of software by the Starlink Project.  %We would like to thank Tom Marsh for the use of his software \emph{molly}, \emph{doppler} and \emph{lcurve}. coauthor now, not neccesary?
The AAVSO International Database was also used in this research, we would like to thank to the worldwide observers from this initiative.  PLP would like to thanks Dr. Phil Carter for his useful Python advice. PLP is supported by a fellowship of the Chilean Government, \emph{``Becas Chile''}, CONICYT.  TRM and DS acknowledge the support of the Science and Technology Facilities Council, grant number ST/L000733/1. We would like to thank the anonymous referee for the useful comments.

\bibliographystyle{mn_new}
\bibliography{mybib}

\end{document}